\documentclass[aps,twocolumn,showpacs,pra,superscriptaddress,floatfix,longbibliography]{revtex4-1}

\usepackage{amsmath,amssymb,amsmath,amsthm,bm,graphicx,xcolor}
\usepackage{natbib,lipsum}
\usepackage{tikz}
\usepackage[breaklinks]{hyperref}
\usepackage{balance}
\usepackage{lastpage}
\hypersetup{colorlinks=true}

\usepackage{color}
\usepackage{tabularx}
\usepackage{epsfig}
\usepackage{graphicx}
\usepackage{wasysym}
\usepackage{dcolumn}
\usepackage{epstopdf}
\usepackage{subfigure}
\usepackage{latexsym}
\usepackage{psfrag}
\usepackage{bbold}

\newcommand{\bra}[1]{\mbox{$\langle #1 |$}}
\newcommand{\bd}[1]{\boldsymbol{ #1 }}
\newcommand{\ket}[1]{\mbox{$| #1 \rangle$}}

\renewcommand{\H}{\mathcal{H}}

\begin{document}

\title{Relating correlation measures: the importance of the energy gap}

\author{Carlos L. Benavides-Riveros}
\email{carlos.benavides-riveros@physik.uni-halle.de}
\affiliation{Institut f\"ur Physik, Martin-Luther-Universit\"at
Halle-Wittenberg, 06120 Halle (Saale), Germany}

\author{Nektarios N. Lathiotakis}
\affiliation{Theoretical and  Physical Chemistry Institute,
National Hellenic Research Foundation, GR-11635 Athens, Greece}

\author{Christian Schilling}
\affiliation{Clarendon Laboratory, University of Oxford, Parks Road, Oxford OX1 3PU, United Kingdom}

\author{Miguel A. L. Marques}
\affiliation{Institut f\"ur Physik, Martin-Luther-Universit\"at
Halle-Wittenberg, 06120 Halle (Saale), Germany}

\date{\today}

\begin{abstract}
The concept of correlation is central to all
approaches that attempt the description of
many-body effects in electronic systems.
Multipartite correlation is a quantum information
theoretical property that is attributed
to quantum states independent of the underlying
physics. In quantum
chemistry, however, the correlation energy (the
energy not seized by the Hartree-Fock ansatz)
plays a more prominent role.
We show that these two different viewpoints on
electron correlation are closely related. The key
ingredient turns out to be the energy gap within
the symmetry-adapted subspace. We then use
a few-site Hubbard model and the stretched H$_2$
to illustrate this connection and to show how the
corresponding measures of correlation compare.
\end{abstract}


% Electron correlation calculations, 31.15.V-
% electronic structure of atoms and molecules, 31.15.xr
%atoms, and molecules
%calculations of, 31.15.-p
%ab initio calculations, 31.15.A-
%alternative approaches to, 31.15.X-
%approximate calculations for, 31.15.B-
%density-functional theory, 31.15.E-
%electron correlation calculations, 31.15.V-
%corrections to, 31.30.-i
%interaction effects on, 31.70.-f
%of magnetic nanoparticles, 75.75.Lf
%theory of, 31.10.+z

\maketitle

\section{Introduction}

Since P.-O.~L\"owdin in the fifties, one
usually defines correlation energy in quantum chemistry
by the difference between the exact ground state (GS)
energy of the system and its Hartree-Fock (HF)
energy~\cite{Lowdin}:
\begin{align}\label{eq:enecorr}
  E_{\rm corr} = E_{\rm GS} - E_{\rm HF}.
\end{align}
Since $E_{\rm HF}$ is an upper bound on $E_{\rm GS}$
the correlation energy is negative by definition.
Beyond HF theory, numerous other methods (such as,
e.g., configuration interaction or coupled-cluster theory)
aim at reconstructing the part of the energy
missing from a single-determinantal description. In fact,
one common indicator of the accuracy of a method is the
percentage of the correlation energy it is able to recover.
Rigorous estimates of the error of the HF energy are
already known for Coulomb systems with large atomic
numbers \cite{bach1992}.

In density-functional theory (DFT), nowadays the workhorse
theory for both quantum chemistry and solid-state physics,
the correlation energy has a slightly different definition.
Instead of HF energy, one can use as reference the energy
obtained by the (exchange only) optimized effective
potential method~\cite{PhysRev.90.317,PhysRevA.14.36,RevModPhys.80.3} which is slightly higher than the HF energy. Clearly, the
choice of the reference energy is arbitrary, as the correlation
energy is not a physical observable. It remains, however, a very
useful tool in understanding and quantifying the magnitude of
many-body effects in given systems.

In recent years a considerable effort has been devoted to
characterize the correlation of a quantum system from a
quantum-information theoretical viewpoint \cite{Horodeckis}.
\emph{A priori}, fermionic correlation is a property of the
many-electron wave function. For the ground state
$\ket{\Psi_{\rm GS}}$, the total correlation can be
quantified by the minimal (Hilbert-Schmidt) distance of
$\ket{\Psi_{\rm GS}}\!\bra{\Psi_{\rm GS}}$
to a single Slater determinant state
\cite{Shimony, Myers2010, metric} or just to
the HF ground state $\ket{\Psi_{\rm HF}}\!\bra{\Psi_{\rm HF}}$,
\begin{eqnarray}\label{eq:over}
D(\Psi_{\rm GS},\Psi_{\rm HF}) &\equiv& \frac{1}{2}
{\rm Tr}\Big[\big(\ket{\Psi_{\rm GS}}\!\bra{\Psi_{\rm GS}}
-\ket{\Psi_{\rm HF}}\!\bra{\Psi_{\rm HF}}\big)^2\Big]\nonumber \\
&=&1-\big|\langle \Psi_{\rm GS}\ket{\Psi_{\rm HF}}\big|^2 .
\end{eqnarray}
This is closely related to the  $L^2$-norm
$\|\Psi_{\rm GS}-\Psi_{\rm HF}\|^2$, that however
is not a good distance measure since it depends on the global
phases of the respective states (which remains a problem
even after restricting to real-valued wave functions). The distance \eqref{eq:over} is bounded between 0 and 1,
reaching the upper value when the overlap between the two
wave functions vanishes. Note that maximising this distance for
fixed $\ket{\Psi_{\rm GS}}$ over all single Slater determinants
is not equivalent to the minimisation of the energy that leads to the
Hartree-Fock orbitals. In fact, such procedure leads to the so-called
Brueckner orbitals~\cite{:/content/aip/journal/jmp/3/6/10.1063/1.1703860,
Zhang},
which are more ``physical'' than Hartree-Fock or Kohn-Sham
orbitals, as they represent much better single-particle
quantities~\cite{Lindgren1976,PhysRevA.31.1273,:/content/aip/journal/jcp/112/16/10.1063/1.481309}.
We note in passing that in DFT it is less common to
measure correlation from the overlap of the wave functions, as the
Kohn-Sham Slater determinant describes a fictitious system and not a
real one.
Further correlation measures involving directly the $N$-fermion
wave function are the Slater rank for two-electron systems
\cite{Cirac, Plastino}, the entanglement classification
for the three-fermion case \cite{Magyares} or the comparison
with uncorrelated states \cite{PhysRevLett.95.123003}.

The nonclassical nature of quantum correlations
and entanglement has enormous implications for quantum
cryptography or quantum computation. Yet, quantifying
correlations and entanglement for systems of identical particles
is a part of an ongoing debate \cite{ReyesLega, Plenio, Bennati, Iemini,
Mikin}. From a practical viewpoint, measuring
correlation is even more challenging for identical particles
since typically only one- and possibly two-particle properties are
experimentally accessible. As a consequence, also simplified
correlation measures involving reduced density operators
were developed. These are, e.g., the squared Frobenius
norm of the cumulant part of the two-particle reduced density
matrix \cite{Mazziotticorr}, the entanglement spectrum and its gap  \cite{Haldane, Thomale}, the von-Neumann entropy
$S(\hat\rho_1) = - \mbox{Tr}[\hat\rho_1 \log{\hat\rho_1}]$ of
the one-particle reduced density operator $\hat\rho_1$
or just the $l^1-$distance $\delta(\vec{n})$ of the
decreasingly-ordered natural occupation numbers
$\vec{n}$ (the eigenvalues of $\hat\rho_1$) to
the ``Hartree-Fock''-point
$\vec{n}_{\rm HF} = (1_1,\ldots,1_N,0_{N+1},\ldots)$ \cite{CS2013}.

A first elementary relation between all those correlation
measures and the concept of correlation energy is obvious:
Each measure attains the minimal value zero whenever
the exact ground state is given by a single Slater
determinant~\cite{Helbig}, i.e.~the correlation energy
vanishes. Furthermore, a monotonous relationship between
the von-Neumann entropy of $\hat{\rho}_1$ and the
density functional definition of correlation energy
has already been observed for some
specific systems~\cite{Smith, Laetitia, Granada}.

In this paper we establish a connection between
those two viewpoints on electron correlation by
providing a concise universal relation between the distance
measure \eqref{eq:over} and the correlation energy
$E_{\rm corr}$. Furthermore, due to the continuity of the
partial trace similar relations between measures
involving reduced density operators and
$E_{\rm corr}$ follow then immediately.

The paper is arranged as follows.
Section \ref{sec:two} presents our main results, while
section \ref{sec_tres} illustrates them for molecular systems. 
The last section provides a conclusion. Technical aspects 
of our work are presented in the appendix.

\section{Main results}
\label{sec:two}

Our starting point is the following theorem.
Let $\hat{H}$ be a Hamiltonian on the Hilbert space
$\H$ with a unique ground state $\ket{\Psi_{\rm GS}}$
and an energy gap $E_{\rm gap} = E_{\rm ES}-E_{\rm GS}$
to the first excited state. Then, for any $\ket{\Psi} \in \H$
with energy $E = \bra{\Psi}\hat{H}\ket{\Psi}$ we have
\begin{align}\label{eq:EvsPsi}
  |\bra{\Psi_{\rm GS}}\Psi\rangle|^2 \geq \frac{E_{\rm ES} - E}
  {E_{\rm gap}}.
\end{align}

The significance of this theorem concerns the case of energy expectation
values $E = \bra{\Psi}\hat{H}\ket{\Psi}$ within the energy
gap $[E_{\rm GS},E_{\rm ES}]$, and relates the energy picture with
the structure of the quantum state. A state $\ket{\Psi}$
has a good overlap with the ground state whenever its energy
expectation value $E$ is close to the ground state energy,
when measured relatively to the energy gap $E_{\rm gap}$.

To prove this theorem we use the spectral decomposition
of $\hat{H} = \sum_{E} E \hat P_E$, where $\hat P_E$ is
the orthogonal projection operator onto the eigenspace of
energy $E$. This yields
\begin{align*}
E = \bra{\Psi}\hat{H}\ket{\Psi} &\geq E_{\rm GS} \bra{\Psi} \hat P_{\rm GS} \ket{\Psi} +E_{\rm ES} \sum_{E\geq E_{\rm ES} } \bra{\Psi}\hat{P}_E\ket{\Psi} \nonumber \\
&= E_{\rm GS} \bra{\Psi} \hat P_{\rm GS} \ket{\Psi} +E_{\rm ES} (1-\bra{\Psi}\hat P_{\rm GS}\ket{\Psi}),
\end{align*}
where we used in the last line
$\sum_{E\geq E_{\rm ES}}\hat P_E=\mathbb{1}-\hat P_{\rm GS}$.
By using $\bra{\Psi}\hat P_{\rm GS}\ket{\Psi}=|\bra{\Psi_{\rm GS}}\Psi\rangle|^2$ this leads to Eq.~(\ref{eq:EvsPsi}) which completes the proof.

From this result, we can deduce that
the distance between the ground state of \emph{any} Hamiltonian
(with a unique ground state) and the corresponding HF ground
state is bounded from above by a function depending on
the energy gap of the system according to
\begin{equation}\label{eq:B1}
  D(\Psi_{\rm GS},\Psi_{\rm HF})
  \leq \frac{|E_{\rm corr}|}{E_{\rm gap}}.
\end{equation}

In practice, the Hamiltonian at hand typically exhibits symmetries.
For instance, the electronic Hamiltonian $\hat{H}$
of atoms and molecules commutes with the total spin. The ground state
inherits this symmetry, i.e.~it lies in an eigenspace
$\mathcal{H}_\sigma = \hat{\pi}_\sigma\mathcal{H}$
of the symmetry operators, where $\hat{\pi}_\sigma$ denotes
the restriction to that subspace with eigenvalue $\sigma$.
Numerical methods are usually adapted to the ground state
symmetry (if possible). A prime example is the restricted HF,
a specific HF ansatz for approximating ground states with
the correct spin symmetries. These considerations on
symmetries allow for a significant improvement of estimate
(\ref{eq:B1}): $\ket{\Psi_{\rm GS}}$ and $\ket{\Psi_{\rm HF}}$
are not only ground state and HF ground state of
$\hat{H}$, respectively, but also of the restricted Hamiltonian
\begin{equation}
\hat{H}_\sigma = \hat{\pi}_\sigma \hat{H} \hat{\pi}_\sigma^\dagger
\end{equation}
acting on the symmetry-adapted Hilbert space
$\mathcal{H}_\sigma$. Application of the estimate (\ref{eq:B1})
to $\hat{H}_\sigma$ implies an improved upper bound:
$E_{\rm gap}$ no longer refers to the gap to the first excited
state but to the first excited state \emph{within} the
symmetry-adapted space $\mathcal{H}_\sigma$ of the
ground state (and may therefore increase considerably).
In the following, $E_{\rm ES}$ will therefore stand for the energy of the first excited state with the same symmetries as the ground state.

The estimate \eqref{eq:B1} is our most significant result.
It establishes a connection between both viewpoints on
electron correlation and shows that the dimensionless
quantity $|E_{\rm corr}| / E_\text{gap}$ provides
a universal upper bound on correlations described by
the wave function. This result also underlines the importance
of the energy gap being the natural reference energy scale. Furthermore,
it is worth noting that estimate (\ref{eq:B1}) implies a similar estimate
for the simplified correlation measure $\delta(\vec{n}) = \mbox{dist}_{l^1}(\vec{n},\vec{n}_{\rm HF})=\sum_{i=1}^N(1-n_i)+\sum_{j > N}n_j$,
since (see Appendix \ref{app:one}):
\begin{equation}\label{eq:dD}
\frac{\delta(\vec{n})}{2N} \leq D(\Psi_{\rm GS},\Psi_{\rm HF}).
\end{equation}
Before we continue a note of caution is in order here. One might
be tempted to apply estimate \eqref{eq:B1} to metals. However,
since metals have a vanishing energy gap and also
$E_{\rm ES} = E_{\rm GS} < E_\text{HF}$,
i.e.~$|E_{\rm corr}| > E_{\rm gap}$, our estimate
has no relevance for them.

To illustrate our results, in the next section we use simple,
analytically solvable systems, namely the two- and
three-site Hubbard model, which are well known for their
capability of exhibiting both, weak and strong (static)
correlation. We study also the stretching of H$_2$, which
is considered a paradigm of the difficulties that
single-determinant methods have with bond dissociation
\cite{Fuchs}.

\section{Numerical investigations}
\label{sec_tres}

\subsection{Hubbard model}

Besides
its importance for solid-state physics, the Hubbard
model is one of the paradigmatic instances
used to simplify the description of strongly correlated
quantum many-body systems.
The Hamiltonian (in second quantization) of the
one-dimensional $r$-site Hubbard model reads:
\begin{align}
  \label{eq:Hamilt}
\hat H = -\frac{t}2 \sum_{i,\sigma}
(c^\dagger_{i\sigma} c_{(i+1)\sigma}
+ h.c. )
+ 2 U \sum_{i} \hat n_{i\uparrow} \hat n_{i \downarrow},
\end{align}
$i \in \{1,2,\dots,r\}$,
where $c^\dagger_{i\sigma}$ and $c_{i\sigma}$ are the fermionic
creation and annihilation operators for a particle on the site $i$
with spin $\sigma \in \{\uparrow, \downarrow \}$ and $\hat n_{i\sigma}
= c^\dagger_{i\sigma} c_{i\sigma}$ is the particle-number operator.
The first term in Eq.~\eqref{eq:Hamilt} describes the hopping between
two neighboring sites while the second represents the on-site interaction.
Periodic boundary conditions in the case $r > 2$ are also assumed.
Achieved experimentally very recently with full control over the quantum state
\cite{PhysRevLett114}, this model may be considered as a simplified tight-binding description of the H$_r$ molecule \cite{Olsen}.

For two fermions on two sites, the eigenstates of $\hat H$
are described by four quantum numbers $\ket{E, s,  m, p}$,
$E$ being the energy,
$(s, m)$ the spin eigenvalues and $p$
the eigenvalue of the operator swapping both sites.
The dimension of the Hilbert space is
$6$, which splits in two parts according to the total
spin: There are three triplet spin states with 0-energy, $\ket{0, 1,
  1, -1}$, $\ket{0, 1, -1, -1}$ and $\ket{0, 1, 0, -1}$, and three
singlets, one of them $\ket{2U, 0, 0, -1}$.
The other two singlets $\ket{E_{\rm GS}, 0, 0, 1}$ and
$\ket{E_{\rm ES}, 0, 0, 1}$ span the spin and translation
symmetry-adapted Hilbert space $\mathcal{H}_{0,0,1}$.
A straightforward computation yields for the ground state
$E_{\rm GS} = U - \sqrt{U^2 + t^2}$
and for the excited state $E_{\rm ES} = U + \sqrt{U^2 + t^2}$.
The restricted HF energy, $E_{\rm HF} = -t + U$, is a reasonable
approximation to the exact ground state energy only for small values of $U/t$.
The unphysical behaviour observed for larger values can be explained
by the contribution of ionic states to the HF wave function \cite{Ziesche}.
The energy gap is given by $2 \sqrt{U^2 + t^2}$.
Since the subspace of $s=m=0$ and $p=+1$ is two-dimensional and since the
restricted HF ground state belongs to it as well, we have
that the equality in \eqref{eq:B1} holds:
$D(\Psi_{\rm GS}, \Psi_{\rm HF})  = |E_{\rm corr}|/E_{\rm gap}$.

For the ground state $\ket{\Psi_{\rm GS}}$, the
corresponding natural occupation numbers follow as
$n_{\pm}(U/t) = (1 \pm 1/\sqrt{1 + U^2/t^2})/2$,
each one with multiplicity two. Note that by defining the
dimensionless energy gap $\Delta = E_{\rm gap}/t$ we
can express the occupation numbers as a function of $\Delta$, leading to
$n_{\pm} (\Delta) = 1/2 \pm 1/\Delta$.
This result shows that the one-particle correlation measures
(von-Neumann entropy and $\delta$-distance) also
depend on the energy gap. In particular, the distance of the natural occupation numbers
to the HF-point follows as $\delta(\Delta) = 2 - 4/\Delta$ which turns out to
saturate the inequality \eqref{eq:dD}.

\begin{figure}[t]
\centering
\includegraphics[width=9cm]{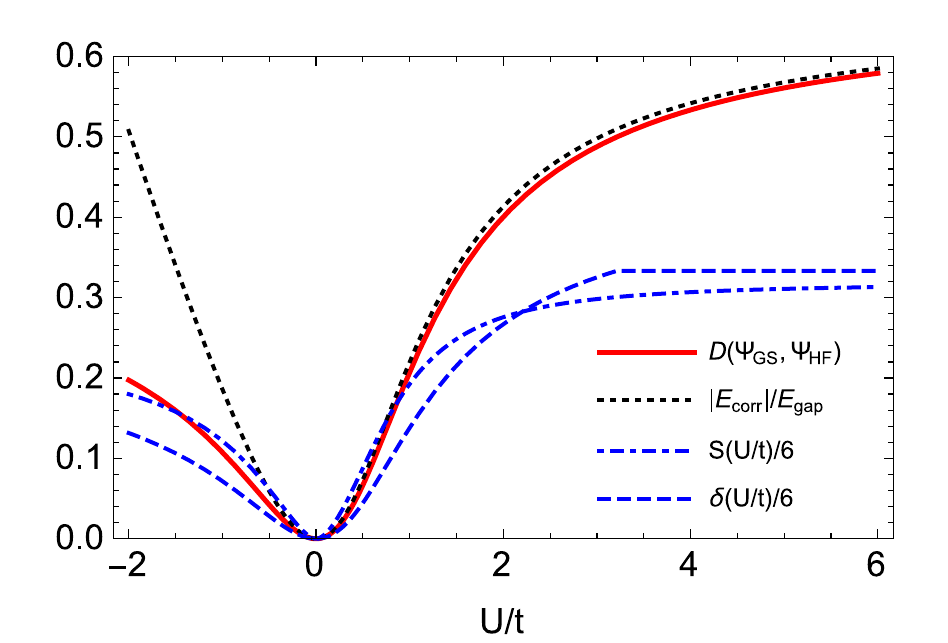}
\caption{For the Hubbard model for three fermions on three sites we present several correlation measures
as functions of the dimensionless coupling $U/t$. These are
  the distance $D (\Psi_{\rm HF}, \Psi_{\rm GS})$ of the GS to the HF state, $|E_{\rm corr}|/E_{\rm gap}$ and the one-particle correlation measures $S$, $\delta$ (see text).}
\label{fig:Hbounds}
\end{figure}

To study the Hubbard model for more than two sites,
we first recall that the Hamiltonian \eqref{eq:Hamilt}
commutes with the total spin vector operator,
its $z$-component and the translation operator
(from the lattice site $i$ to the next site $i+1$),
with eigenvalues $e^{i2\pi p/r}$ with $p\in\{0,1,\dots,r-1\}$.
The Hamiltonian is block diagonal with respect to those
symmetries (see Appendix \ref{app:two}).
For the case of three fermions on three sites,
the spectrum of the Hubbard model restricted
to the subspace that corresponds to $s = \scriptstyle\frac12$,
$m = \scriptstyle\frac12$ and $p = 2$ is
given by \cite{CS2015Hubbard}:
$$
E_{j}(U,t) = -2\sqrt{Q}\cos\bigg(\frac{\theta-2\pi j}3\bigg) + \frac{4U}3,
\quad j \in \{0,1,2\},
$$
where $Q = 28 U^2/9 + 3 t^2/4$
and $\cos\theta = 8 U^3/(27Q^{3/2})$. The dimensionless
energy gap is
$\Delta(U/t) = (E_{\rm ES} - E_{\rm GS})/t = -2\sqrt{3Q} \sin[(\theta-\pi)/3]/t$.
For positive values of the dimensionless coupling $U/t$,
$\Delta(U/t) =  3/2 + 4(U/t)^2/9 + \mathcal{O}((U/t)^3)$.
For negative values, the energy gap is bounded from above:
$\Delta(U/t) \rightarrow 1.73205$.

In Fig.~\ref{fig:Hbounds} we plot several correlation
measures as a function of $U/t$ for this model.
As expected, all curves increase monotonically with
the strength of the interaction. For the positive region
$U/t \geq 0$, the curve for $|E_{\rm corr}|/E_{\rm gap}$
follows very closely the one for $D(\Psi_{\rm GS},\Psi_{\rm HF})$
confirming the significance of our estimate (\ref{eq:B1}).
Both curves converge to the same value ($2/3$) for
$U/t \rightarrow \infty$ . However,
for negative values of $U/t$ the estimate
loses its significance. This is based on the fact
that a significant part of the weight of the HF ground
state lies on \emph{higher} excited states. In addition,
the energy gap is getting of the same order of magnitude
as the correlation energy, leading to a rapid growth of our
bound. In the strong correlation regime, beyond $U/t < -3.375$,
$|E_{\rm corr}| > E_{\rm gap}$ and
our estimate has no significance. For positive values of $U/t$
the energy gap increases monotonously.
Note that the quantity $|E_{\rm corr}|/E_{\rm gap}$ provides
a much better estimate on the quantum state overlap (\ref{eq:over}) than the
von-Neumann entropy or the $l^1-$distance to the HF-point.
The latter ones (the blue curves in Fig.~\ref{fig:Hbounds})
saturate very soon in contrast to the red and black ones.
This shows the limitation of the one-particle picture to
measure total fermion correlation.

\begin{figure}[t]
\centering
\includegraphics[width=9cm]{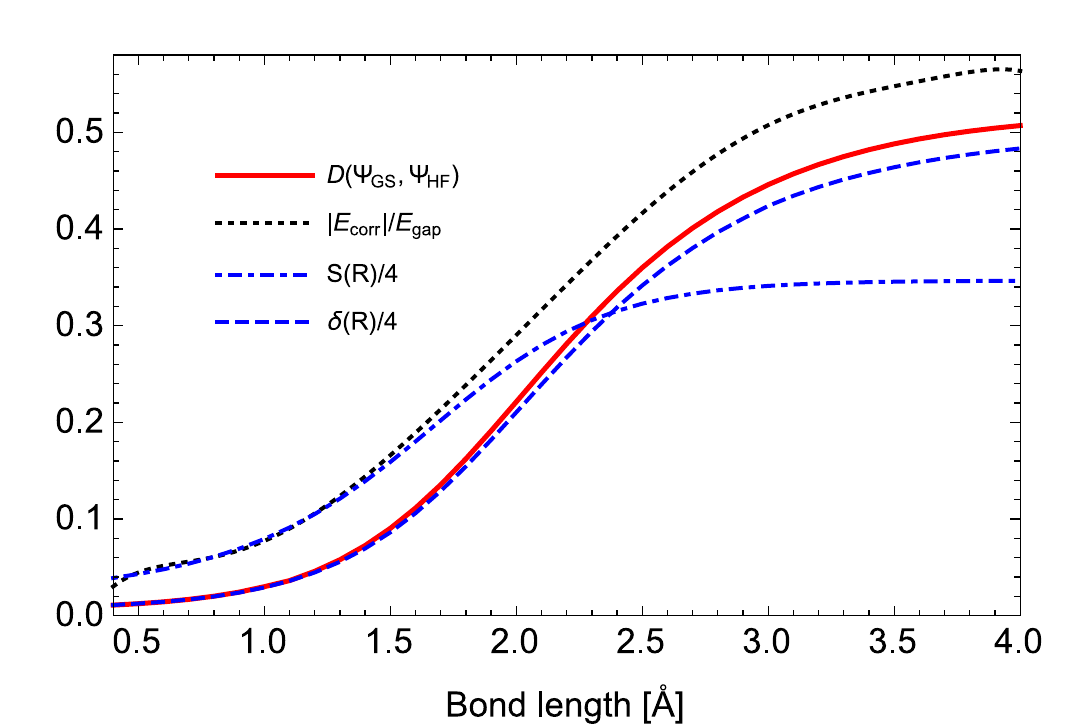}
\caption{For the stretched H$_2$ we present several
correlation measures as functions of the bond length.
These are the distance $D (\Psi_{\rm HF}, \Psi_{\rm GS})$
of the GS to the HF state, $|E_{\rm corr}|/E_{\rm gap}$
and the one-particle correlation measures $S$, $\delta$ (see text).}
\label{fig:H2energies}
\end{figure}

\subsection{ The stretched H$_2$}

As a second example we look at the archetypal
instance of strong (static) correlation, i.e.~the stretched
dihydrogen H$_2$~\cite{CoulsonFish}, which we
analyze numerically using a
cc-pVTZ basis set. In its dissociation limit,
this system is commonly used as a benchmark to produce
exchange-correlation functionals for strong static correlations~\cite{PhysRevLett.87.133004,Matito2016}.
The HF approach describes well the
equilibrium chemical bond, but fails dramatically
as the molecule is stretched. It is also known
that DFT functionals describe the covalent bond well,
but the predicted energy is overestimated in the dissociation
limit due to delocalization, static-correlation
and self-interaction errors \cite{Cohen792}.
Around the equilibrium separation (0.74 \AA), electronic
correlation is not particularly large and the HF state therefore
approximates significantly well the ground-state wave function.
The first excited state of H$_2$ with the same
symmetry of the ground state ($s = m = 0$)
is the second excited state. Around the equilibrium
geometry, the energy gap diminishes as the interatomic
distance is elongated. As for the two-site Hubbard model,
close to equilibrium, the bound $|E_{\rm corr}|/E_{\rm gap}$
provides a good estimate on the correlation measure $D(\Psi_{\rm GS},\Psi_{\rm HF})$.
Remarkably, as shown in Fig.~\ref{fig:H2energies},
beyond the equilibrium bound, where the static correlation
effects can be observed, $|E_{\rm corr}|/E_{\rm gap}$
reproduces the behaviour of the distance $D(\Psi_{\rm GS},\Psi_{\rm HF})$.
The same holds for the $\delta$-distance, which
is largely due to the fact that for two-fermion models the
value of the first occupation number is approximately
the square of the projection of the ground state onto the
HF configuration. In contrast, the von-Neumann entropy
saturates very soon.

\section{Conclusion}

In conclusion, we have connected both viewpoints on fermion correlation
by providing the universal estimate (\ref{eq:B1}). It connects the
measure of total fermion correlation (as property that can be
attributed to quantum states independent of the underlying
physics) and the correlation energy
(commonly used in quantum chemistry). The quantity that
connects both measures is the energy gap of the corresponding
block Hamiltonian with the \emph{same} symmetry as the ground
state. Moreover, due to the continuity of the partial trace,
similar estimates follow for several correlation measures
resorting to reduced-particle information only. Yet, as it can be
inferred from their early saturation shown in Fig.~\ref{fig:Hbounds},
the significance of such simplified correlation measures is limited.

Since the quantity $|E_{\rm corr}|/E_{\rm gap}$ provides an
estimate on the overlap between the HF and the exact
ground state wave function our work may allow
one to use the sophisticated concept of multipartite entanglement
developed and explored in quantum information theory for a more
systematic study of strongly correlated systems. In particular,
our work suggests an additional tool for
describing the possible failure of DFT in reconstructing specific
properties of a given quantum system. This failure can be either
attributed to a rather poor reconstruction of the systems ground
state energy or to the failure of the effective  method (e.g.~Kohn-Sham)
 in reconstructing many-particle properties from one-particle
information. The latter case would be reflected by poor saturation
 of the inequality (\ref{eq:B1}) while the first one corresponds to
 a large correlation energy (requiring a multi-reference method
 instead \cite{Cohen792, Caruso, CSHFZPC, CBRV}).

\section*{Acknowledgements}
We thank D.\hspace{0.5mm}Gross and
M.\hspace{0.5mm}Springborg for helpful
discussions. C.L.B.R. thanks the Clarendon
Laboratory at the University of Oxford for the
warm hospitality. We acknowledge financial
support from the Hellenic Ministry of Education
(through ESPA) and from the GSRT through
``Advanced Materials and Devices'' program
(MIS:5002409) (N.N.L.), the Oxford Martin
Programme on Bio-Inspired Quantum Technologies,
the UK Engineering, Physical Sciences Research
Council (Grant EP/P007155/1) (C.S.) and the
DFG through projects SFB-762 and MA 6787/1-1 (M.A.L.M.).

\appendix

\section{Proof of estimate \eqref{eq:dD}}
\label{app:one}

We consider an $N$-fermion Hilbert space $\mathcal{H}_{N}^{(f)}$ where the underlying one-particle Hilbert space $\mathcal{H}_1^{(d)}$ has dimension $d \in \mathbb{N}\cup\{\infty\}$. For $\ket{\Psi}\in \mathcal{H}_{N}^{(f)}$ we can determine its one-particle reduced density operator $\hat{\rho}_1$
(trace-normalized to $N$) and the vector  $\vec{\lambda} = (\lambda_i)_{i=1}^d$ of decreasingly-ordered eigenvalues of $\hat{\rho}_1$ (natural occupation numbers). Let $\{\ket{\chi_j}\}_{j=1}^d$ be a Brueckner orthonormal basis for $\mathcal{H}_{1}^{(d)}$, i.e.~the specific Slater determinant $\ket{\bd{\chi}}=\ket{\chi_1,\ldots,\chi_N}$ maximizes the overlap with $\ket{\Psi}$. Furthermore we introduce $\delta(\vec{x})= \sum_{i=1}^N (1-x_i)+\sum_{j=N+1}^d x_j$ and $\hat{n}_i$ as the particle number operator for state $\ket{\chi_i}$. Obviously, $\delta(\vec{\lambda})$ is the $l^1$-distance of $\vec{\lambda}$ to the ``Hartree-Fock''-point $(1,\ldots,1,0,\ldots)$. In the following we prove the estimate
\begin{equation}\label{eq:dvsSD}
\frac{\delta(\vec{\lambda})}{2\,\mbox{min}(N,d-N)} \leq 1-|\langle \bd{\chi}\ket{\Psi}|^2\,.
\end{equation}
For this, we introduce the particle number expectation values $n_i = \bra{\Psi}\hat{n}_i\ket{\Psi}$ and $\hat{\delta}= \delta((\hat{n}_i)_{i=1}^d)$. Since the spectrum of $\hat{\delta}$ is given by $\{0,1,\ldots,2M\}$ with $M=\mbox{min}(N,d-N)$ we find
\begin{align}
\delta(\vec{n}) &= \bra{\Psi}\hat{\delta}\ket{\Psi} =\sum_{d=1}^{2M} d \,\|\hat{P}_d \Psi\|_{L^2}^2 \leq 2M\sum_{d=1}^{2M} \|\hat{P}_d \Psi\|_{L^2}^2
\nonumber
\\ &= 2M\,\big(1-\|\hat{P}_0 \Psi\|_{L^2}^2\big)\,,
\end{align}
where we have used the spectral decomposition $\hat{\delta}=\oplus_{d=0}^{2M} \,d \,\hat{P}_d$ of $\hat{\delta}$. By using $\hat{P}_0=\ket{\bd{\chi}}\!\bra{\bd{\chi}}$ and the fact that the vector $\vec{\lambda}$ of decreasingly-ordered eigenvalues of $\hat{\rho}_1$ majorizes any other vector of occupation numbers (particularly $\vec{n}$) we obtain
\begin{equation}
 \delta(\vec{\lambda}) \leq \delta(\vec{n}) \leq 2M \big(1-|\langle \bd{\chi}\ket{\Psi}|^2\big)\,.
\end{equation}
Since $\ket{\bd{\chi}}$ maximizes the overlap with $\ket{\Psi}$, we eventually find for the Hartree-Fock ground state $\ket{\Psi_{\rm HF}}$ (or any other single Slater determinant)
\begin{equation}
\frac{\delta(\vec{\lambda})}{2M} \leq 1-|\langle \Psi_{\rm HF} \ket{\Psi}|^2\,,
\end{equation}
i.e.~estimate \eqref{eq:dD}.

\section{Analytic solution of the Hubbard model for three electrons on three sites}
\label{app:two}

In this section we recall the analytical solution of the three-site
Hubbard model for three electrons which was already presented in Ref.~\cite{CS2015Hubbard}.
For the Hubbard model, the one-body reduced density matrix
is diagonal in the basis of the Bloch orbitals, which satisfy
$\hat T_1 \ket{q} = e^{i\varphi q}\ket{q}$, where $\varphi = 2\pi/r$,
$\hat T_1$ is the 1-particle translation operator and $\hat T = \bigotimes^N_{i=1} \hat T_1$. The creation operators in the
Bloch basis set read:
$\tilde c^\dagger_{q\sigma} = \frac1{\sqrt{r}} \sum_{k=1}^{r}
e^{i\varphi q k} c^\dagger_{k\sigma}$,
$q\in \{0,1,\dots,r-1\}$. To block diagonalize this Hamiltonian one
can employ the natural-orbital basis set generated by $\{\ket{q}\}$
and then split the total Hilbert space with respect to the spin quantum
numbers $s$ and $m$. For example, the only state with maximal magnetic
number $m = r/2$ is $\tilde c^\dagger_{0\uparrow} \tilde
c^\dagger_{1\uparrow} \cdots \tilde c^\dagger_{r-1\uparrow} \ket{\rm vac}$,
which spans the one-dimensional subspace $\H_{\scriptstyle\frac{r}2,
\scriptstyle\frac{r}2,\eta}$ ($\eta = 0$ for $r$ odd or $\eta = r/2$ otherwise),
as defined by the direct sum of the total Hilbert space:
\begin{align}
\label{eq:hS}
\H =
\bigoplus^{N/2}_{s=s_{-}}\bigoplus^{s}_{m=-s} \bigoplus^{r-1}_{p=0} \H_{s,m,p}.
\end{align}

\begin{figure}[t]
\centering
\includegraphics[width=9cm]{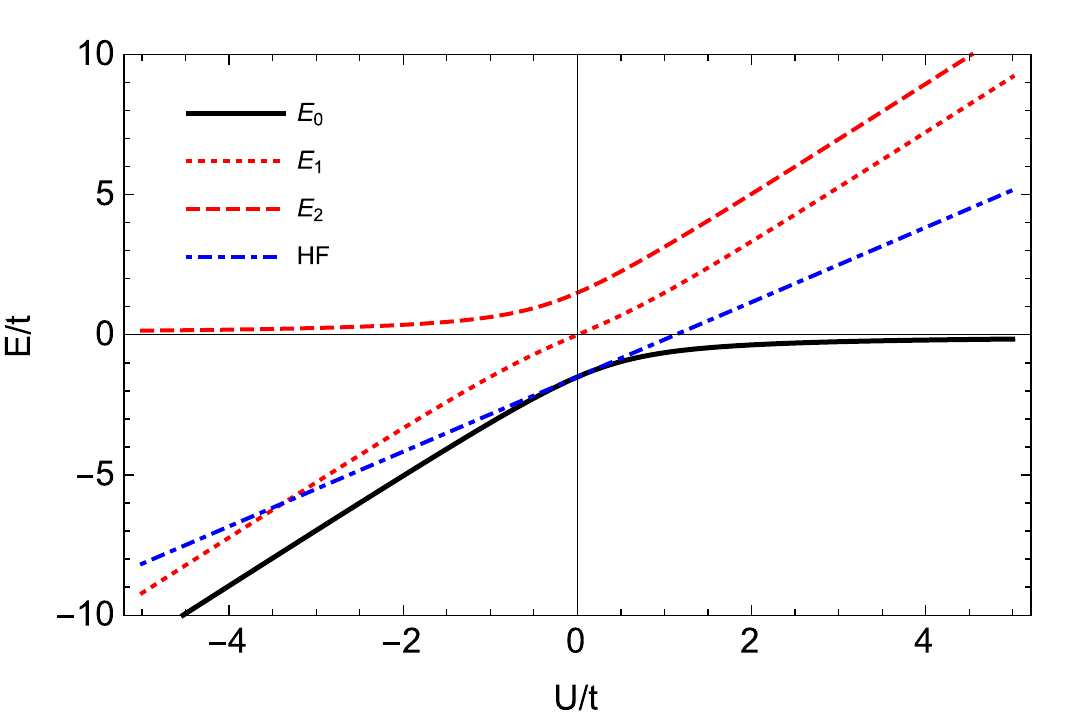}
\caption{Energy spectrum of the three-site three-fermion Hubbard model
  restricted to the Hilbert subspace where the ground state lies.
  The Hartree-Fock energy is also shown.}
\label{fig:bounds}
\end{figure}

For the case of three fermions on three sites, the total spin quantum number
$s$ can take only two values $\tfrac32$ and $\tfrac12$.
For $s=  \tfrac12$, thanks to the fact that the Hamiltonian is invariant
under simultaneous flipping of all spins, the results for $m= -\tfrac12$
are identical to the case $m= \tfrac12$. The latter is related to the eight
dimensional Hilbert space $\H_{\scriptstyle\frac12,\scriptstyle\frac12,2}\oplus
\H_{\scriptstyle\frac12,\scriptstyle\frac12,1}\oplus\H_{\scriptstyle\frac12,\scriptstyle\frac12,0}$, where
\begin{eqnarray*}
&\H_{\scriptstyle\frac12,\scriptstyle\frac12,2} = {\rm span} \{\ket{0\uparrow0\downarrow2\uparrow},
\ket{2\uparrow2\downarrow1\uparrow},
 \ket{1\uparrow1\downarrow0\uparrow}
\}, \\
&\H_{\scriptstyle\frac12,\scriptstyle\frac12,1} = {\rm span}
\{\ket{0\uparrow0\downarrow1\uparrow},
 \ket{2\uparrow2\downarrow0\uparrow},\ket{1\uparrow1\downarrow2\uparrow}\}.
\end{eqnarray*}
The translation invariant subspace $\H_{\scriptstyle\frac12,\scriptstyle\frac12,0}$
is two dimensional and can be built with two spin-compensated linear combinations of
the following three configurations:
$\ket{0\uparrow1\uparrow2\downarrow}$,
$\ket{0\uparrow1\downarrow2\uparrow}$ and
$\ket{0\downarrow1\uparrow2\uparrow}$.
As an elementary exercise one verifies that the Hamiltonian restricted
to each one of the subspaces
$\H_{\scriptstyle\frac12,\scriptstyle\frac12,2}$ and
$\H_{\scriptstyle\frac12,\scriptstyle\frac12,1}$
leads to the same $3\times 3$ matrix. Indeed,
\begin{align}
\label{eq:3time3}
\hat H|_{\H_{\scriptstyle\frac12,\scriptstyle\frac12,1}} =
\hat H|_{\H_{\scriptstyle\frac12,\scriptstyle\frac12,2}}.
\end{align}
It is worth noting that the same configuration appears
in the description of the spin-compensated Lithium
isoelectronic series~\cite{BenavLiQuasi}.
Moreover, since the diagonalization of any of the
Hamiltonians \eqref{eq:3time3}
can be performed  analytically, an expression for the
energy spectrum can be exactly known
 \cite{CS2015Hubbard}:
$$
E_{j} = -2\sqrt{Q}\cos\bigg(\frac{\theta-2\pi j}3\bigg) + \frac{4U}3,
$$
for $j = 0, 1, 2$. Here $Q = 28 U^2/9 + 3 t^2/4$
and $\cos\theta = 8 U^3/(27Q^{3/2})$. See Fig.~\ref{fig:bounds}.
 The energy gap $E_{\rm gap}$ is then given by
$E_{1} - E_{0}$.

\balance

\bibliography{correlation_PRA_v2}

%merlin.mbs aipnum4-1.bst 2010-07-25 4.21a (PWD, AO, DPC) hacked
%Control: key (0)
%Control: author (8) initials jnrlst
%Control: editor formatted (1) identically to author
%Control: production of article title (-1) disabled
%Control: page (0) single
%Control: year (1) truncated
%Control: production of eprint (0) enabled
\begin{thebibliography}{44}%
\makeatletter
\providecommand \@ifxundefined [1]{%
 \@ifx{#1\undefined}
}%
\providecommand \@ifnum [1]{%
 \ifnum #1\expandafter \@firstoftwo
 \else \expandafter \@secondoftwo
 \fi
}%
\providecommand \@ifx [1]{%
 \ifx #1\expandafter \@firstoftwo
 \else \expandafter \@secondoftwo
 \fi
}%
\providecommand \natexlab [1]{#1}%
\providecommand \enquote  [1]{``#1''}%
\providecommand \bibnamefont  [1]{#1}%
\providecommand \bibfnamefont [1]{#1}%
\providecommand \citenamefont [1]{#1}%
\providecommand \href@noop [0]{\@secondoftwo}%
\providecommand \href [0]{\begingroup \@sanitize@url \@href}%
\providecommand \@href[1]{\@@startlink{#1}\@@href}%
\providecommand \@@href[1]{\endgroup#1\@@endlink}%
\providecommand \@sanitize@url [0]{\catcode `\\12\catcode `\$12\catcode
  `\&12\catcode `\#12\catcode `\^12\catcode `\_12\catcode `\%12\relax}%
\providecommand \@@startlink[1]{}%
\providecommand \@@endlink[0]{}%
\providecommand \url  [0]{\begingroup\@sanitize@url \@url }%
\providecommand \@url [1]{\endgroup\@href {#1}{\urlprefix }}%
\providecommand \urlprefix  [0]{URL }%
\providecommand \Eprint [0]{\href }%
\providecommand \doibase [0]{http://dx.doi.org/}%
\providecommand \selectlanguage [0]{\@gobble}%
\providecommand \bibinfo  [0]{\@secondoftwo}%
\providecommand \bibfield  [0]{\@secondoftwo}%
\providecommand \translation [1]{[#1]}%
\providecommand \BibitemOpen [0]{}%
\providecommand \bibitemStop [0]{}%
\providecommand \bibitemNoStop [0]{.\EOS\space}%
\providecommand \EOS [0]{\spacefactor3000\relax}%
\providecommand \BibitemShut  [1]{\csname bibitem#1\endcsname}%
\let\auto@bib@innerbib\@empty
%</preamble>
\bibitem [{\citenamefont {L\"owdin}(1955)}]{Lowdin}%
  \BibitemOpen
  \bibfield  {author} {\bibinfo {author} {\bibfnamefont {P.-O.}\ \bibnamefont
  {L\"owdin}},\ }\href {\doibase 10.1103/PhysRev.97.1509} {\bibfield  {journal}
  {\bibinfo  {journal} {Phys. Rev.}\ }\textbf {\bibinfo {volume} {97}},\
  \bibinfo {pages} {1509} (\bibinfo {year} {1955})}\BibitemShut {NoStop}%
\bibitem [{\citenamefont {Bach}(1992)}]{bach1992}%
  \BibitemOpen
  \bibfield  {author} {\bibinfo {author} {\bibfnamefont {V.}~\bibnamefont
  {Bach}},\ }\href {\doibase 10.1007/BF02097241} {\bibfield  {journal}
  {\bibinfo  {journal} {Commun. Math. Phys.}\ }\textbf {\bibinfo {volume}
  {147}},\ \bibinfo {pages} {527} (\bibinfo {year} {1992})}\BibitemShut
  {NoStop}%
\bibitem [{\citenamefont {Sharp}\ and\ \citenamefont
  {Horton}(1953)}]{PhysRev.90.317}%
  \BibitemOpen
  \bibfield  {author} {\bibinfo {author} {\bibfnamefont {R.~T.}\ \bibnamefont
  {Sharp}}\ and\ \bibinfo {author} {\bibfnamefont {G.~K.}\ \bibnamefont
  {Horton}},\ }\href {\doibase 10.1103/PhysRev.90.317} {\bibfield  {journal}
  {\bibinfo  {journal} {Phys. Rev.}\ }\textbf {\bibinfo {volume} {90}},\
  \bibinfo {pages} {317} (\bibinfo {year} {1953})}\BibitemShut {NoStop}%
\bibitem [{\citenamefont {Talman}\ and\ \citenamefont
  {Shadwick}(1976)}]{PhysRevA.14.36}%
  \BibitemOpen
  \bibfield  {author} {\bibinfo {author} {\bibfnamefont {J.~D.}\ \bibnamefont
  {Talman}}\ and\ \bibinfo {author} {\bibfnamefont {W.~F.}\ \bibnamefont
  {Shadwick}},\ }\href {\doibase 10.1103/PhysRevA.14.36} {\bibfield  {journal}
  {\bibinfo  {journal} {Phys. Rev. A}\ }\textbf {\bibinfo {volume} {14}},\
  \bibinfo {pages} {36} (\bibinfo {year} {1976})}\BibitemShut {NoStop}%
\bibitem [{\citenamefont {K\"ummel}\ and\ \citenamefont
  {Kronik}(2008)}]{RevModPhys.80.3}%
  \BibitemOpen
  \bibfield  {author} {\bibinfo {author} {\bibfnamefont {S.}~\bibnamefont
  {K\"ummel}}\ and\ \bibinfo {author} {\bibfnamefont {L.}~\bibnamefont
  {Kronik}},\ }\href {\doibase 10.1103/RevModPhys.80.3} {\bibfield  {journal}
  {\bibinfo  {journal} {Rev. Mod. Phys.}\ }\textbf {\bibinfo {volume} {80}},\
  \bibinfo {pages} {3} (\bibinfo {year} {2008})}\BibitemShut {NoStop}%
\bibitem [{\citenamefont {Horodecki}\ \emph {et~al.}(2009)\citenamefont
  {Horodecki}, \citenamefont {Horodecki}, \citenamefont {Horodecki},\ and\
  \citenamefont {Horodecki}}]{Horodeckis}%
  \BibitemOpen
  \bibfield  {author} {\bibinfo {author} {\bibfnamefont {R.}~\bibnamefont
  {Horodecki}}, \bibinfo {author} {\bibfnamefont {P.}~\bibnamefont
  {Horodecki}}, \bibinfo {author} {\bibfnamefont {M.}~\bibnamefont
  {Horodecki}}, \ and\ \bibinfo {author} {\bibfnamefont {K.}~\bibnamefont
  {Horodecki}},\ }\href {\doibase 10.1103/RevModPhys.81.865} {\bibfield
  {journal} {\bibinfo  {journal} {Rev. Mod. Phys.}\ }\textbf {\bibinfo {volume}
  {81}},\ \bibinfo {pages} {865} (\bibinfo {year} {2009})}\BibitemShut
  {NoStop}%
\bibitem [{\citenamefont {Shimony}(1995)}]{Shimony}%
  \BibitemOpen
  \bibfield  {author} {\bibinfo {author} {\bibfnamefont {A.}~\bibnamefont
  {Shimony}},\ }\href {\doibase 10.1111/j.1749-6632.1995.tb39008.x} {\bibfield
  {journal} {\bibinfo  {journal} {Ann. N. Y. Acad. Sci.}\ }\textbf {\bibinfo
  {volume} {755}},\ \bibinfo {pages} {675} (\bibinfo {year}
  {1995})}\BibitemShut {NoStop}%
\bibitem [{\citenamefont {Myers}\ and\ \citenamefont {Wu}(2010)}]{Myers2010}%
  \BibitemOpen
  \bibfield  {author} {\bibinfo {author} {\bibfnamefont {J.~M.}\ \bibnamefont
  {Myers}}\ and\ \bibinfo {author} {\bibfnamefont {T.~T.}\ \bibnamefont {Wu}},\
  }\href {\doibase 10.1007/s11128-009-0146-5} {\bibfield  {journal} {\bibinfo
  {journal} {Quantum Inf. Process.}\ }\textbf {\bibinfo {volume} {9}},\
  \bibinfo {pages} {239} (\bibinfo {year} {2010})}\BibitemShut {NoStop}%
\bibitem [{\citenamefont {D'Amico}\ \emph {et~al.}(2011)\citenamefont
  {D'Amico}, \citenamefont {Coe}, \citenamefont {Fran\c{c}a},\ and\
  \citenamefont {Capelle}}]{metric}%
  \BibitemOpen
  \bibfield  {author} {\bibinfo {author} {\bibfnamefont {I.}~\bibnamefont
  {D'Amico}}, \bibinfo {author} {\bibfnamefont {J.~P.}\ \bibnamefont {Coe}},
  \bibinfo {author} {\bibfnamefont {V.~V.}\ \bibnamefont {Fran\c{c}a}}, \ and\
  \bibinfo {author} {\bibfnamefont {K.}~\bibnamefont {Capelle}},\ }\href
  {\doibase 10.1103/PhysRevLett.106.050401} {\bibfield  {journal} {\bibinfo
  {journal} {Phys. Rev. Lett.}\ }\textbf {\bibinfo {volume} {106}},\ \bibinfo
  {pages} {050401} (\bibinfo {year} {2011})}\BibitemShut {NoStop}%
\bibitem [{\citenamefont
  {L{\"o}wdin}(1962)}]{:/content/aip/journal/jmp/3/6/10.1063/1.1703860}%
  \BibitemOpen
  \bibfield  {author} {\bibinfo {author} {\bibfnamefont {P.}~\bibnamefont
  {L{\"o}wdin}},\ }\href
  {http://scitation.aip.org/content/aip/journal/jmp/3/6/10.1063/1.1703860}
  {\bibfield  {journal} {\bibinfo  {journal} {J. Math. Phys.}\ }\textbf
  {\bibinfo {volume} {3}},\ \bibinfo {pages} {1171} (\bibinfo {year}
  {1962})}\BibitemShut {NoStop}%
\bibitem [{\citenamefont {Zhang}\ and\ \citenamefont {Mauser}(2016)}]{Zhang}%
  \BibitemOpen
  \bibfield  {author} {\bibinfo {author} {\bibfnamefont {J.~M.}\ \bibnamefont
  {Zhang}}\ and\ \bibinfo {author} {\bibfnamefont {N.~J.}\ \bibnamefont
  {Mauser}},\ }\href {\doibase 10.1103/PhysRevA.94.032513} {\bibfield
  {journal} {\bibinfo  {journal} {Phys. Rev. A}\ }\textbf {\bibinfo {volume}
  {94}},\ \bibinfo {pages} {032513} (\bibinfo {year} {2016})}\BibitemShut
  {NoStop}%
\bibitem [{\citenamefont {Lindgren}, \citenamefont {Lindgren},\ and\
  \citenamefont {M{\aa}rtensson}(1976)}]{Lindgren1976}%
  \BibitemOpen
  \bibfield  {author} {\bibinfo {author} {\bibfnamefont {I.}~\bibnamefont
  {Lindgren}}, \bibinfo {author} {\bibfnamefont {J.}~\bibnamefont {Lindgren}},
  \ and\ \bibinfo {author} {\bibfnamefont {A.-M.}\ \bibnamefont
  {M{\aa}rtensson}},\ }\href {\doibase 10.1007/BF01437866} {\bibfield
  {journal} {\bibinfo  {journal} {Z. Phys. A}\ }\textbf {\bibinfo {volume}
  {279}},\ \bibinfo {pages} {113} (\bibinfo {year} {1976})}\BibitemShut
  {NoStop}%
\bibitem [{\citenamefont {Lindgren}(1985)}]{PhysRevA.31.1273}%
  \BibitemOpen
  \bibfield  {author} {\bibinfo {author} {\bibfnamefont {I.}~\bibnamefont
  {Lindgren}},\ }\href {\doibase 10.1103/PhysRevA.31.1273} {\bibfield
  {journal} {\bibinfo  {journal} {Phys. Rev. A}\ }\textbf {\bibinfo {volume}
  {31}},\ \bibinfo {pages} {1273} (\bibinfo {year} {1985})}\BibitemShut
  {NoStop}%
\bibitem [{\citenamefont {He{\ss}elmann}\ and\ \citenamefont
  {Jansen}(2000)}]{:/content/aip/journal/jcp/112/16/10.1063/1.481309}%
  \BibitemOpen
  \bibfield  {author} {\bibinfo {author} {\bibfnamefont {A.}~\bibnamefont
  {He{\ss}elmann}}\ and\ \bibinfo {author} {\bibfnamefont {G.}~\bibnamefont
  {Jansen}},\ }\href
  {http://scitation.aip.org/content/aip/journal/jcp/112/16/10.1063/1.481309}
  {\bibfield  {journal} {\bibinfo  {journal} {J. Chem. Phys.}\ }\textbf
  {\bibinfo {volume} {112}},\ \bibinfo {pages} {6949} (\bibinfo {year}
  {2000})}\BibitemShut {NoStop}%
\bibitem [{\citenamefont {Schliemann}\ \emph {et~al.}(2001)\citenamefont
  {Schliemann}, \citenamefont {Cirac}, \citenamefont {Ku\ifmmode~\acute{s}\else
  \'{s}\fi{}}, \citenamefont {Lewenstein},\ and\ \citenamefont {Loss}}]{Cirac}%
  \BibitemOpen
  \bibfield  {author} {\bibinfo {author} {\bibfnamefont {J.}~\bibnamefont
  {Schliemann}}, \bibinfo {author} {\bibfnamefont {J.~I.}\ \bibnamefont
  {Cirac}}, \bibinfo {author} {\bibfnamefont {M.}~\bibnamefont
  {Ku\ifmmode~\acute{s}\else \'{s}\fi{}}}, \bibinfo {author} {\bibfnamefont
  {M.}~\bibnamefont {Lewenstein}}, \ and\ \bibinfo {author} {\bibfnamefont
  {D.}~\bibnamefont {Loss}},\ }\href {\doibase 10.1103/PhysRevA.64.022303}
  {\bibfield  {journal} {\bibinfo  {journal} {Phys. Rev. A}\ }\textbf {\bibinfo
  {volume} {64}},\ \bibinfo {pages} {022303} (\bibinfo {year}
  {2001})}\BibitemShut {NoStop}%
\bibitem [{\citenamefont {Plastino}, \citenamefont {Manzano},\ and\
  \citenamefont {Dehesa}(2009)}]{Plastino}%
  \BibitemOpen
  \bibfield  {author} {\bibinfo {author} {\bibfnamefont {A.~R.}\ \bibnamefont
  {Plastino}}, \bibinfo {author} {\bibfnamefont {D.}~\bibnamefont {Manzano}}, \
  and\ \bibinfo {author} {\bibfnamefont {J.~S.}\ \bibnamefont {Dehesa}},\
  }\href {http://stacks.iop.org/0295-5075/86/i=2/a=20005} {\bibfield  {journal}
  {\bibinfo  {journal} {EPL}\ }\textbf {\bibinfo {volume} {86}},\ \bibinfo
  {pages} {20005} (\bibinfo {year} {2009})}\BibitemShut {NoStop}%
\bibitem [{\citenamefont {S\'arosi}\ and\ \citenamefont
  {L\'evay}(2014)}]{Magyares}%
  \BibitemOpen
  \bibfield  {author} {\bibinfo {author} {\bibfnamefont {G.}~\bibnamefont
  {S\'arosi}}\ and\ \bibinfo {author} {\bibfnamefont {P.}~\bibnamefont
  {L\'evay}},\ }\href {\doibase 10.1103/PhysRevA.89.042310} {\bibfield
  {journal} {\bibinfo  {journal} {Phys. Rev. A}\ }\textbf {\bibinfo {volume}
  {89}},\ \bibinfo {pages} {042310} (\bibinfo {year} {2014})}\BibitemShut
  {NoStop}%
\bibitem [{\citenamefont {Gottlieb}\ and\ \citenamefont
  {Mauser}(2005)}]{PhysRevLett.95.123003}%
  \BibitemOpen
  \bibfield  {author} {\bibinfo {author} {\bibfnamefont {A.~D.}\ \bibnamefont
  {Gottlieb}}\ and\ \bibinfo {author} {\bibfnamefont {N.~J.}\ \bibnamefont
  {Mauser}},\ }\href {\doibase 10.1103/PhysRevLett.95.123003} {\bibfield
  {journal} {\bibinfo  {journal} {Phys. Rev. Lett.}\ }\textbf {\bibinfo
  {volume} {95}},\ \bibinfo {pages} {123003} (\bibinfo {year}
  {2005})}\BibitemShut {NoStop}%
\bibitem [{\citenamefont {Balachandran}\ \emph {et~al.}(2013)\citenamefont
  {Balachandran}, \citenamefont {Govindarajan}, \citenamefont {de~Queiroz},\
  and\ \citenamefont {Reyes-Lega}}]{ReyesLega}%
  \BibitemOpen
  \bibfield  {author} {\bibinfo {author} {\bibfnamefont {A.~P.}\ \bibnamefont
  {Balachandran}}, \bibinfo {author} {\bibfnamefont {T.~R.}\ \bibnamefont
  {Govindarajan}}, \bibinfo {author} {\bibfnamefont {A.~R.}\ \bibnamefont
  {de~Queiroz}}, \ and\ \bibinfo {author} {\bibfnamefont {A.~F.}\ \bibnamefont
  {Reyes-Lega}},\ }\href {\doibase 10.1103/PhysRevLett.110.080503} {\bibfield
  {journal} {\bibinfo  {journal} {Phys. Rev. Lett.}\ }\textbf {\bibinfo
  {volume} {110}},\ \bibinfo {pages} {080503} (\bibinfo {year}
  {2013})}\BibitemShut {NoStop}%
\bibitem [{\citenamefont {Killoran}, \citenamefont {Cramer},\ and\
  \citenamefont {Plenio}(2014)}]{Plenio}%
  \BibitemOpen
  \bibfield  {author} {\bibinfo {author} {\bibfnamefont {N.}~\bibnamefont
  {Killoran}}, \bibinfo {author} {\bibfnamefont {M.}~\bibnamefont {Cramer}}, \
  and\ \bibinfo {author} {\bibfnamefont {M.~B.}\ \bibnamefont {Plenio}},\
  }\href {\doibase 10.1103/PhysRevLett.112.150501} {\bibfield  {journal}
  {\bibinfo  {journal} {Phys. Rev. Lett.}\ }\textbf {\bibinfo {volume} {112}},\
  \bibinfo {pages} {150501} (\bibinfo {year} {2014})}\BibitemShut {NoStop}%
\bibitem [{\citenamefont {Benatti}, \citenamefont {Alipour},\ and\
  \citenamefont {Rezakhani}(2014)}]{Bennati}%
  \BibitemOpen
  \bibfield  {author} {\bibinfo {author} {\bibfnamefont {F.}~\bibnamefont
  {Benatti}}, \bibinfo {author} {\bibfnamefont {S.}~\bibnamefont {Alipour}}, \
  and\ \bibinfo {author} {\bibfnamefont {A.~T.}\ \bibnamefont {Rezakhani}},\
  }\href {http://stacks.iop.org/1367-2630/16/i=1/a=015023} {\bibfield
  {journal} {\bibinfo  {journal} {New J. Phys.}\ }\textbf {\bibinfo {volume}
  {16}},\ \bibinfo {pages} {015023} (\bibinfo {year} {2014})}\BibitemShut
  {NoStop}%
\bibitem [{\citenamefont {Iemini}, \citenamefont {Maciel},\ and\ \citenamefont
  {Vianna}(2015)}]{Iemini}%
  \BibitemOpen
  \bibfield  {author} {\bibinfo {author} {\bibfnamefont {F.}~\bibnamefont
  {Iemini}}, \bibinfo {author} {\bibfnamefont {T.~O.}\ \bibnamefont {Maciel}},
  \ and\ \bibinfo {author} {\bibfnamefont {R.~O.}\ \bibnamefont {Vianna}},\
  }\href {\doibase 10.1103/PhysRevB.92.075423} {\bibfield  {journal} {\bibinfo
  {journal} {Phys. Rev. B}\ }\textbf {\bibinfo {volume} {92}},\ \bibinfo
  {pages} {075423} (\bibinfo {year} {2015})}\BibitemShut {NoStop}%
\bibitem [{\citenamefont {Miklin}, \citenamefont {Moroder},\ and\ \citenamefont
  {G\"uhne}(2016)}]{Mikin}%
  \BibitemOpen
  \bibfield  {author} {\bibinfo {author} {\bibfnamefont {N.}~\bibnamefont
  {Miklin}}, \bibinfo {author} {\bibfnamefont {T.}~\bibnamefont {Moroder}}, \
  and\ \bibinfo {author} {\bibfnamefont {O.}~\bibnamefont {G\"uhne}},\ }\href
  {\doibase 10.1103/PhysRevA.93.020104} {\bibfield  {journal} {\bibinfo
  {journal} {Phys. Rev. A}\ }\textbf {\bibinfo {volume} {93}},\ \bibinfo
  {pages} {020104} (\bibinfo {year} {2016})}\BibitemShut {NoStop}%
\bibitem [{\citenamefont {Juhász}\ and\ \citenamefont
  {Mazziotti}(2006)}]{Mazziotticorr}%
  \BibitemOpen
  \bibfield  {author} {\bibinfo {author} {\bibfnamefont {T.}~\bibnamefont
  {Juhász}}\ and\ \bibinfo {author} {\bibfnamefont {D.~A.}\ \bibnamefont
  {Mazziotti}},\ }\href
  {http://scitation.aip.org/content/aip/journal/jcp/125/17/10.1063/1.2378768}
  {\bibfield  {journal} {\bibinfo  {journal} {J. Chem. Phys.}\ }\textbf
  {\bibinfo {volume} {125}},\ \bibinfo {eid} {174105} (\bibinfo {year}
  {2006})}\BibitemShut {NoStop}%
\bibitem [{\citenamefont {Li}\ and\ \citenamefont {Haldane}(2008)}]{Haldane}%
  \BibitemOpen
  \bibfield  {author} {\bibinfo {author} {\bibfnamefont {H.}~\bibnamefont
  {Li}}\ and\ \bibinfo {author} {\bibfnamefont {F.~D.~M.}\ \bibnamefont
  {Haldane}},\ }\href {\doibase 10.1103/PhysRevLett.101.010504} {\bibfield
  {journal} {\bibinfo  {journal} {Phys. Rev. Lett.}\ }\textbf {\bibinfo
  {volume} {101}},\ \bibinfo {pages} {010504} (\bibinfo {year}
  {2008})}\BibitemShut {NoStop}%
\bibitem [{\citenamefont {Thomale}\ \emph {et~al.}(2010)\citenamefont
  {Thomale}, \citenamefont {Sterdyniak}, \citenamefont {Regnault},\ and\
  \citenamefont {Bernevig}}]{Thomale}%
  \BibitemOpen
  \bibfield  {author} {\bibinfo {author} {\bibfnamefont {R.}~\bibnamefont
  {Thomale}}, \bibinfo {author} {\bibfnamefont {A.}~\bibnamefont {Sterdyniak}},
  \bibinfo {author} {\bibfnamefont {N.}~\bibnamefont {Regnault}}, \ and\
  \bibinfo {author} {\bibfnamefont {B.~A.}\ \bibnamefont {Bernevig}},\ }\href
  {\doibase 10.1103/PhysRevLett.104.180502} {\bibfield  {journal} {\bibinfo
  {journal} {Phys. Rev. Lett.}\ }\textbf {\bibinfo {volume} {104}},\ \bibinfo
  {pages} {180502} (\bibinfo {year} {2010})}\BibitemShut {NoStop}%
\bibitem [{\citenamefont {Schilling}, \citenamefont {Gross},\ and\
  \citenamefont {Christandl}(2013)}]{CS2013}%
  \BibitemOpen
  \bibfield  {author} {\bibinfo {author} {\bibfnamefont {C.}~\bibnamefont
  {Schilling}}, \bibinfo {author} {\bibfnamefont {D.}~\bibnamefont {Gross}}, \
  and\ \bibinfo {author} {\bibfnamefont {M.}~\bibnamefont {Christandl}},\
  }\href {\doibase 10.1103/PhysRevLett.110.040404} {\bibfield  {journal}
  {\bibinfo  {journal} {Phys. Rev. Lett.}\ }\textbf {\bibinfo {volume} {110}},\
  \bibinfo {pages} {040404} (\bibinfo {year} {2013})}\BibitemShut {NoStop}%
\bibitem [{\citenamefont {Helbig}, \citenamefont {Tokatly},\ and\ \citenamefont
  {Rubio}(2010)}]{Helbig}%
  \BibitemOpen
  \bibfield  {author} {\bibinfo {author} {\bibfnamefont {N.}~\bibnamefont
  {Helbig}}, \bibinfo {author} {\bibfnamefont {I.~V.}\ \bibnamefont {Tokatly}},
  \ and\ \bibinfo {author} {\bibfnamefont {A.}~\bibnamefont {Rubio}},\ }\href
  {\doibase 10.1103/PhysRevA.81.022504} {\bibfield  {journal} {\bibinfo
  {journal} {Phys. Rev. A}\ }\textbf {\bibinfo {volume} {81}},\ \bibinfo
  {pages} {022504} (\bibinfo {year} {2010})}\BibitemShut {NoStop}%
\bibitem [{\citenamefont {Smith}, \citenamefont {Schmider},\ and\ \citenamefont
  {Smith}(2002)}]{Smith}%
  \BibitemOpen
  \bibfield  {author} {\bibinfo {author} {\bibfnamefont {G.~T.}\ \bibnamefont
  {Smith}}, \bibinfo {author} {\bibfnamefont {H.~L.}\ \bibnamefont {Schmider}},
  \ and\ \bibinfo {author} {\bibfnamefont {V.~H.}\ \bibnamefont {Smith}},\
  }\href {\doibase 10.1103/PhysRevA.65.032508} {\bibfield  {journal} {\bibinfo
  {journal} {Phys. Rev. A}\ }\textbf {\bibinfo {volume} {65}},\ \bibinfo
  {pages} {032508} (\bibinfo {year} {2002})}\BibitemShut {NoStop}%
\bibitem [{\citenamefont {Benavides-Riveros}, \citenamefont
  {Gracia-Bond\'{\i}a},\ and\ \citenamefont {V\'arilly}(2012)}]{Laetitia}%
  \BibitemOpen
  \bibfield  {author} {\bibinfo {author} {\bibfnamefont {C.~L.}\ \bibnamefont
  {Benavides-Riveros}}, \bibinfo {author} {\bibfnamefont {J.~M.}\ \bibnamefont
  {Gracia-Bond\'{\i}a}}, \ and\ \bibinfo {author} {\bibfnamefont {J.~C.}\
  \bibnamefont {V\'arilly}},\ }\href {\doibase 10.1103/PhysRevA.86.022525}
  {\bibfield  {journal} {\bibinfo  {journal} {Phys. Rev. A}\ }\textbf {\bibinfo
  {volume} {86}},\ \bibinfo {pages} {022525} (\bibinfo {year}
  {2012})}\BibitemShut {NoStop}%
\bibitem [{\citenamefont {Benavides-Riveros}, \citenamefont {Toranzo},\ and\
  \citenamefont {Dehesa}(2014)}]{Granada}%
  \BibitemOpen
  \bibfield  {author} {\bibinfo {author} {\bibfnamefont {C.~L.}\ \bibnamefont
  {Benavides-Riveros}}, \bibinfo {author} {\bibfnamefont {I.~V.}\ \bibnamefont
  {Toranzo}}, \ and\ \bibinfo {author} {\bibfnamefont {J.~S.}\ \bibnamefont
  {Dehesa}},\ }\href {http://stacks.iop.org/0953-4075/47/i=19/a=195503}
  {\bibfield  {journal} {\bibinfo  {journal} {J. Phys. B: At. Mol. Opt. Phys.}\
  }\textbf {\bibinfo {volume} {47}},\ \bibinfo {pages} {195503} (\bibinfo
  {year} {2014})}\BibitemShut {NoStop}%
\bibitem [{\citenamefont {Fuchs}\ \emph {et~al.}(2005)\citenamefont {Fuchs},
  \citenamefont {Niquet}, \citenamefont {Gonze},\ and\ \citenamefont
  {Burke}}]{Fuchs}%
  \BibitemOpen
  \bibfield  {author} {\bibinfo {author} {\bibfnamefont {M.}~\bibnamefont
  {Fuchs}}, \bibinfo {author} {\bibfnamefont {Y.-M.}\ \bibnamefont {Niquet}},
  \bibinfo {author} {\bibfnamefont {X.}~\bibnamefont {Gonze}}, \ and\ \bibinfo
  {author} {\bibfnamefont {K.}~\bibnamefont {Burke}},\ }\href
  {http://dx.doi.org/10.1063/1.1858371} {\bibfield  {journal} {\bibinfo
  {journal} {J. Chem. Phys.}\ }\textbf {\bibinfo {volume} {122}},\ \bibinfo
  {pages} {094116} (\bibinfo {year} {2005})}\BibitemShut {NoStop}%
\bibitem [{\citenamefont {Murmann}\ \emph {et~al.}(2015)\citenamefont
  {Murmann}, \citenamefont {Bergschneider}, \citenamefont {Klinkhamer},
  \citenamefont {Z\"urn}, \citenamefont {Lompe},\ and\ \citenamefont
  {Jochim}}]{PhysRevLett114}%
  \BibitemOpen
  \bibfield  {author} {\bibinfo {author} {\bibfnamefont {S.}~\bibnamefont
  {Murmann}}, \bibinfo {author} {\bibfnamefont {A.}~\bibnamefont
  {Bergschneider}}, \bibinfo {author} {\bibfnamefont {V.~M.}\ \bibnamefont
  {Klinkhamer}}, \bibinfo {author} {\bibfnamefont {G.}~\bibnamefont {Z\"urn}},
  \bibinfo {author} {\bibfnamefont {T.}~\bibnamefont {Lompe}}, \ and\ \bibinfo
  {author} {\bibfnamefont {S.}~\bibnamefont {Jochim}},\ }\href {\doibase
  10.1103/PhysRevLett.114.080402} {\bibfield  {journal} {\bibinfo  {journal}
  {Phys. Rev. Lett.}\ }\textbf {\bibinfo {volume} {114}},\ \bibinfo {pages}
  {080402} (\bibinfo {year} {2015})}\BibitemShut {NoStop}%
\bibitem [{\citenamefont {Olsen}\ and\ \citenamefont {Thygesen}(2014)}]{Olsen}%
  \BibitemOpen
  \bibfield  {author} {\bibinfo {author} {\bibfnamefont {T.}~\bibnamefont
  {Olsen}}\ and\ \bibinfo {author} {\bibfnamefont {K.~S.}\ \bibnamefont
  {Thygesen}},\ }\href
  {http://scitation.aip.org/content/aip/journal/jcp/140/16/10.1063/1.4871875}
  {\bibfield  {journal} {\bibinfo  {journal} {J. Chem. Phys.}\ }\textbf
  {\bibinfo {volume} {140}},\ \bibinfo {eid} {164116} (\bibinfo {year}
  {2014})}\BibitemShut {NoStop}%
\bibitem [{\citenamefont {Ziesche}\ \emph {et~al.}(1997)\citenamefont
  {Ziesche}, \citenamefont {Gunnarsson}, \citenamefont {John},\ and\
  \citenamefont {Beck}}]{Ziesche}%
  \BibitemOpen
  \bibfield  {author} {\bibinfo {author} {\bibfnamefont {P.}~\bibnamefont
  {Ziesche}}, \bibinfo {author} {\bibfnamefont {O.}~\bibnamefont {Gunnarsson}},
  \bibinfo {author} {\bibfnamefont {W.}~\bibnamefont {John}}, \ and\ \bibinfo
  {author} {\bibfnamefont {H.}~\bibnamefont {Beck}},\ }\href {\doibase
  10.1103/PhysRevB.55.10270} {\bibfield  {journal} {\bibinfo  {journal} {Phys.
  Rev. B}\ }\textbf {\bibinfo {volume} {55}},\ \bibinfo {pages} {10270}
  (\bibinfo {year} {1997})}\BibitemShut {NoStop}%
\bibitem [{\citenamefont {Schilling}(2015)}]{CS2015Hubbard}%
  \BibitemOpen
  \bibfield  {author} {\bibinfo {author} {\bibfnamefont {C.}~\bibnamefont
  {Schilling}},\ }\href {\doibase 10.1103/PhysRevB.92.155149} {\bibfield
  {journal} {\bibinfo  {journal} {Phys. Rev. B}\ }\textbf {\bibinfo {volume}
  {92}},\ \bibinfo {pages} {155149} (\bibinfo {year} {2015})}\BibitemShut
  {NoStop}%
\bibitem [{\citenamefont {Coulson}\ and\ \citenamefont
  {Fischer}(1949)}]{CoulsonFish}%
  \BibitemOpen
  \bibfield  {author} {\bibinfo {author} {\bibfnamefont {C.~A.}\ \bibnamefont
  {Coulson}}\ and\ \bibinfo {author} {\bibfnamefont {I.}~\bibnamefont
  {Fischer}},\ }\href {\doibase 10.1080/14786444908521726} {\bibfield
  {journal} {\bibinfo  {journal} {Philos. Mag.}\ }\textbf {\bibinfo {volume}
  {40}},\ \bibinfo {pages} {386} (\bibinfo {year} {1949})}\BibitemShut
  {NoStop}%
\bibitem [{\citenamefont {Baerends}(2001)}]{PhysRevLett.87.133004}%
  \BibitemOpen
  \bibfield  {author} {\bibinfo {author} {\bibfnamefont {E.~J.}\ \bibnamefont
  {Baerends}},\ }\href {\doibase 10.1103/PhysRevLett.87.133004} {\bibfield
  {journal} {\bibinfo  {journal} {Phys. Rev. Lett.}\ }\textbf {\bibinfo
  {volume} {87}},\ \bibinfo {pages} {133004} (\bibinfo {year}
  {2001})}\BibitemShut {NoStop}%
\bibitem [{\citenamefont {Matito}\ \emph {et~al.}(2016)\citenamefont {Matito},
  \citenamefont {Casanova}, \citenamefont {Lopez},\ and\ \citenamefont
  {Ugalde}}]{Matito2016}%
  \BibitemOpen
  \bibfield  {author} {\bibinfo {author} {\bibfnamefont {E.}~\bibnamefont
  {Matito}}, \bibinfo {author} {\bibfnamefont {D.}~\bibnamefont {Casanova}},
  \bibinfo {author} {\bibfnamefont {X.}~\bibnamefont {Lopez}}, \ and\ \bibinfo
  {author} {\bibfnamefont {J.~M.}\ \bibnamefont {Ugalde}},\ }\href {\doibase
  10.1007/s00214-016-1982-x} {\bibfield  {journal} {\bibinfo  {journal} {Theor.
  Chem. Acc.}\ }\textbf {\bibinfo {volume} {135}},\ \bibinfo {pages} {226}
  (\bibinfo {year} {2016})}\BibitemShut {NoStop}%
\bibitem [{\citenamefont {Cohen}, \citenamefont {Mori-S{\'a}nchez},\ and\
  \citenamefont {Yang}(2008)}]{Cohen792}%
  \BibitemOpen
  \bibfield  {author} {\bibinfo {author} {\bibfnamefont {A.~J.}\ \bibnamefont
  {Cohen}}, \bibinfo {author} {\bibfnamefont {P.}~\bibnamefont
  {Mori-S{\'a}nchez}}, \ and\ \bibinfo {author} {\bibfnamefont
  {W.}~\bibnamefont {Yang}},\ }\href {\doibase 10.1126/science.1158722}
  {\bibfield  {journal} {\bibinfo  {journal} {Science}\ }\textbf {\bibinfo
  {volume} {321}},\ \bibinfo {pages} {792} (\bibinfo {year}
  {2008})}\BibitemShut {NoStop}%
\bibitem [{\citenamefont {Hellgren}\ \emph {et~al.}(2015)\citenamefont
  {Hellgren}, \citenamefont {Caruso}, \citenamefont {Rohr}, \citenamefont
  {Ren}, \citenamefont {Rubio}, \citenamefont {Scheffler},\ and\ \citenamefont
  {Rinke}}]{Caruso}%
  \BibitemOpen
  \bibfield  {author} {\bibinfo {author} {\bibfnamefont {M.}~\bibnamefont
  {Hellgren}}, \bibinfo {author} {\bibfnamefont {F.}~\bibnamefont {Caruso}},
  \bibinfo {author} {\bibfnamefont {D.~R.}\ \bibnamefont {Rohr}}, \bibinfo
  {author} {\bibfnamefont {X.}~\bibnamefont {Ren}}, \bibinfo {author}
  {\bibfnamefont {A.}~\bibnamefont {Rubio}}, \bibinfo {author} {\bibfnamefont
  {M.}~\bibnamefont {Scheffler}}, \ and\ \bibinfo {author} {\bibfnamefont
  {P.}~\bibnamefont {Rinke}},\ }\href {\doibase 10.1103/PhysRevB.91.165110}
  {\bibfield  {journal} {\bibinfo  {journal} {Phys. Rev. B}\ }\textbf {\bibinfo
  {volume} {91}},\ \bibinfo {pages} {165110} (\bibinfo {year}
  {2015})}\BibitemShut {NoStop}%
\bibitem [{\citenamefont {Benavides-Riveros}\ and\ \citenamefont
  {Schilling}(2016)}]{CSHFZPC}%
  \BibitemOpen
  \bibfield  {author} {\bibinfo {author} {\bibfnamefont {C.~L.}\ \bibnamefont
  {Benavides-Riveros}}\ and\ \bibinfo {author} {\bibfnamefont {C.}~\bibnamefont
  {Schilling}},\ }\href
  {http://www.degruyter.com/view/j/zpch.2016.230.issue-5-7/zpch-2015-0732/zpch-2015-0732.xml?format=INT}
  {\bibfield  {journal} {\bibinfo  {journal} {Z. Phys. Chem.}\ }\textbf
  {\bibinfo {volume} {230}},\ \bibinfo {pages} {703} (\bibinfo {year}
  {2016})}\BibitemShut {NoStop}%
\bibitem [{\citenamefont {Schilling}, \citenamefont {Benavides-Riveros},\ and\
  \citenamefont {Vrana}(2017)}]{CBRV}%
  \BibitemOpen
  \bibfield  {author} {\bibinfo {author} {\bibfnamefont {C.}~\bibnamefont
  {Schilling}}, \bibinfo {author} {\bibfnamefont {C.~L.}\ \bibnamefont
  {Benavides-Riveros}}, \ and\ \bibinfo {author} {\bibfnamefont
  {P.}~\bibnamefont {Vrana}},\ }\href {https://arxiv.org/abs/1703.01612}
  {\bibfield  {journal} {\bibinfo  {journal} {arXiv:1703.01612}\ } (\bibinfo
  {year} {2017})}\BibitemShut {NoStop}%
\bibitem [{\citenamefont {Benavides-Riveros}, \citenamefont {Gracia-Bondia},\
  and\ \citenamefont {Spring\-borg}(2013)}]{BenavLiQuasi}%
  \BibitemOpen
  \bibfield  {author} {\bibinfo {author} {\bibfnamefont {C.~L.}\ \bibnamefont
  {Benavides-Riveros}}, \bibinfo {author} {\bibfnamefont {J.~M.}\ \bibnamefont
  {Gracia-Bondia}}, \ and\ \bibinfo {author} {\bibfnamefont {M.}~\bibnamefont
  {Spring\-borg}},\ }\href {\doibase 10.1103/PhysRevA.88.022508} {\bibfield
  {journal} {\bibinfo  {journal} {Phys. Rev. A}\ }\textbf {\bibinfo {volume}
  {88}},\ \bibinfo {pages} {022508} (\bibinfo {year} {2013})}\BibitemShut
  {NoStop}%
\end{thebibliography}%

\end{document}